\documentclass[twocolumn,aps,superscriptaddress,showpacs]{revtex4}
\usepackage{amssymb}
\usepackage{amsmath}
\usepackage{mathrsfs}
\usepackage{graphicx}
\usepackage{subfigure}
\usepackage[normalem]{ulem}
\usepackage[dvips]{color}
\usepackage{academicons} 
\usepackage{xcolor}
\usepackage{hyperref}

\hypersetup{hypertex=true,
colorlinks=true,
linkcolor=magenta,
citecolor=magenta,
urlcolor=cyan} 
\definecolor{orcidgreen}{HTML}{A6CE39}
\newcommand{\orcid}[1]{\href{https://orcid.org/#1}{\textcolor{orcidgreen}{\aiOrcid}}}
\begin{document}
\title{Fluctuation of Temperature in the Polyakov-loop extended Nambu--Jona-Lasinio Model}
\author{He Liu \orcid{0000-0002-8658-2851}}
\email{liuhe@qut.edu.cn}
\affiliation{Science School, Qingdao University of Technology, Qingdao 266520, China}
\affiliation{The Research Center of Theoretical Physics, Qingdao University of Technology, Qingdao 266033, China}
\author{Peng Wu}
\affiliation{Science School, Qingdao University of Technology, Qingdao 266520, China}
\affiliation{The Research Center of Theoretical Physics, Qingdao University of Technology, Qingdao 266033, China}
\author{Hong-Ming Liu \orcid{0000-0002-6479-9785}}
\email{liuhongming13@126.com}
\affiliation{Science School, Qingdao University of Technology, Qingdao 266520, China}
\affiliation{The Research Center of Theoretical Physics, Qingdao University of Technology, Qingdao 266033, China}
\author{Peng-Cheng Chu \orcid{0000-0001-7311-9684}}
\email{kyois@126.com}
\affiliation{Science School, Qingdao University of Technology, Qingdao 266520, China}
\affiliation{The Research Center of Theoretical Physics, Qingdao University of Technology, Qingdao 266033, China}
\date{\today}
\begin{abstract}
We investigate temperature fluctuations in hot QCD matter using a 3-flavor Polyakov-loop extended Nambu--Jona-Lasinio (PNJL) model. The high-order cumulant ratios $R_{n2}$ ($n>2$) exhibit non-monotonic variations across the chiral phase transition, characterized by slight fluctuations in the chiral crossover region and significant oscillations around the critical point. In contrast, distinct peak and dip structures are observed in the cumulant ratios at low baryon chemical potential. These structures gradually weaken and eventually vanish at high chemical potential as they compete with the sharpening of the chiral phase transition, particularly near the critical point and the first-order phase transition. Our results indicate that these non-monotonic peak and dip structures in high-order cumulant ratios are associated with the deconfinement phase transition. This study quantitatively analyzes temperature fluctuation behavior across different phase transition regions, and the findings are expected to be observed and validated in heavy-ion collision experiments through measurements of event-by-event mean transverse momentum fluctuations.
\end{abstract}

\maketitle

\section{Introduction}
Investigating the phase structure of Quantum Chromodynamics (QCD) is one of the central goals of heavy-ion collision experiments. It is of great significance for revealing the fundamental properties and evolution of strongly interacting matter under extreme conditions, and for understanding the state of matter in compact stars and the early universe. First-principles calculations based on lattice QCD indicate that near zero baryon chemical potential ($\mu_B$), the phase transition between the quark-gluon plasma (QGP) and the hadron resonance gas (HRG) is a smooth crossover~\cite{Aok06,Gup11,Bor14}. Studies using effective models such as the Nambu--Jona-Lasinio (NJL) model, as well as advanced functional methods including the Dyson-Schwinger equation (DSE) and the functional renormalization group (fRG)~\cite{Sch99,Zhu00,Fu08,Fuk08,Qin11,Sch12,Che16,Liu16,Fu20,Liu21}, suggest that at large $\mu_B$ this transition may become first-order, connected to the crossover region via a critical endpoint (CEP). In relativistic heavy-ion collision experiments conducted at facilities such as the Relativistic Heavy Ion Collider (RHIC) and the Large Hadron Collider (LHC), the creation of QGP offers a unique opportunity to explore the fundamental properties of QCD under extreme conditions~\cite{Shu80,Cor04,Ada05,Adc05,Gar20}. To search for the first-order phase transition and the possible CEP of QCD matter, the Beam Energy Scan (BES-I and BES-II) programs have been carried out at RHIC by varying the collision energy. Event-by-event net-proton number fluctuations in nuclear collision experiments are considered a sensitive probe for locating the QCD critical endpoint~\cite{Hei01,Jeo00,Vol99,Asa00}. Recently, the non-monotonic dependence of net-proton fluctuations, which are sensitive to the correlation length, on collision energy has indeed been observed in the precision BES-II data from the STAR Collaboration, particularly the dip structure around 19.6 GeV~\cite{Ada21L,Abd21,Abd22,Abo25}. Simultaneously, the STAR experiment has measured the energy dependence of other observables sensitive to the CEP and/or the first-order phase transition, including baryon directed flow~\cite{Ada14L,Ada18}, pion HBT radii~\cite{Ada15,Ada21C}, intermittency of charged hadrons~\cite{Abd23B}, and the light nuclei yield ratio ($N_t \times N_p / N_d^2$)~\cite{Abd23L}. Non-monotonic energy dependencies have been observed in all these observables, with peak or dip structures appearing roughly in the energy range $\sqrt{s_{NN}} \approx 7.7-39$ GeV. These intriguing findings have sparked widespread interest, and more accurate BES-II measurements in the near future will provide further insight into the QCD phase diagram.

Compared to net-baryon number fluctuations, temperature fluctuations have also gained attention in recent years, offering a potentially powerful probe for studying QCD thermodynamics and phase transitions. Advances in heavy-ion collision experiments now allow the isolation of thermal fluctuations from confounding effects such as initial-state geometric fluctuations, flow contributions, and other non-thermal sources~\cite{Gar12,Sch14,Aad24,Zha25}, enabling the extraction of temperature fluctuations from event-by-event mean transverse momentum fluctuations of final-state charged particles~\cite{Gav04}. In hydrodynamic simulations, temperature varies with time and spatial coordinates. Specifically, it decreases over time as the system expands into the vacuum. For the equation of state, a simple proportionality has been observed in hydrodynamic simulations between the average transverse momentum $\langle p_T\rangle$ of final-state particles and the effective temperature $T_{\text{eff}}$~\cite{Gar20}. Here $T_{\text{eff}}$ always lies between the final and initial temperatures and roughly corresponds to the average temperature at a time scale on the order of the nuclear radius.As one of the fundamental properties of matter, the speed of sound ($c_s$) can convey QCD phase structure information~\cite{Sor21,He22,Liu24,Yan24}. The mean transverse momentum of final-state particles has been used to determine $c_s$ through the relation $c_s^2 = dp/d\varepsilon \sim d\ln\langle p_T\rangle / d\ln N_{ch}$. Recently, the CMS Collaboration applied this method to measure the speed of sound at $T_{\text{eff}} = 219 \pm 8$ MeV, obtaining $c_s^2 = 0.241 \pm 0.016$, in perfect agreement with lattice QCD calculations~\cite{Hay24}.  Prior to this, the string percolation model had also been widely used to predict several observables in the high energy physics for both theoretical and experimental results~\cite{Bra15}. In the string percolation framework, the fluctuations of the string tension give rise to the thermal distribution of the transverse momentum with a temperature which can be interpreted as the temperature of the initial state of the system. Taking the experimentally determined chemical freeze-out temperature  $T_c = 167.7\pm 2.6$ MeV as the phase transition temperature and using the equation $T(\rho_c)=\sqrt{\langle p_T \rangle^2/(2F_{TL}(\rho_c))}$ with the percolation threshold  $\rho_c = 1.2$ and $F_{TL}(\rho_c) = 0.66$, one obtains  $\sqrt{\langle p_T \rangle^2} = 207.2\pm 3.3$ MeV, which is close to the value of  200 MeV commonly adopted in phenomenological applications~\cite{Ada03}.  A comprehensive review covering the applications of the string percolation model, including shear viscosity, the trace anomaly, and sound velocity, can be found in Refs.~\cite{Sch11}. Meanwhile, a new study provides additional constraints on the equation of state of hot QCD matter and demonstrates that the covariance of the average transverse momentum at two rapidities, $\text{Cov}_{\langle p_T\rangle}(\eta_1,\eta_2)$, and its associated decorrelation measures, $R_{p_T}(\eta_1,\eta_2)$ and $r_{p_T}(\eta, \eta_{\text{ref}})$, exhibit strong sensitivity to the stiffness of the QGP equation of state~\cite{Liu25}. Consequently, event-by-event mean transverse momentum fluctuations have been extensively measured across various collision energies and systems at different heavy-ion facilities, opening a new avenue for exploring the QCD phase diagram~\cite{App99,Ada03,Adl04,Ach24}.

Building on this foundation, temperature fluctuation signatures may provide insights into the phase transition dynamics across the QCD phase diagram. In this work, we employ a 3-flavor Polyakov-loop extended Nambu--Jona-Lasinio (PNJL) model to systematically analyze temperature fluctuations, revealing non-monotonic variations across the chiral phase transition and deconfinement phase transition. Importantly, this study quantitatively analyzes temperature fluctuation behavior across different phase transition regions, and the findings are expected to be observed and validated in heavy-ion collision experiments through measurements of event-by-event mean transverse momentum fluctuations.

\section{Theoretical model and methods}
The thermodynamic potential density of the three-flavor Polyakov-loop extended Nambu--Jona-Lasinio (PNJL) model at finite temperature $T$ is given by
\begin{eqnarray}\label{eq1}
\Omega_{\textrm{PNJL}} &=& \mathcal{U}(\Phi,\bar{\Phi},T)+G_S(\sigma_u^2+\sigma_d^2+\sigma_s^2)-4K\sigma_u\sigma_d\sigma_s
\notag\\
&-&2N_c\sum_{i=u,d,s}\int_0^\Lambda\frac{d^3p}{(2\pi)^3}E_i
\notag\\
&-&2T\sum_{i=u,d,s}\int\frac{d^3p}{(2\pi)^3}(F_1+F_2),
\end{eqnarray}
where $F_1=\ln[1+3\Phi e^{-\beta(E_i-\mu_i)}+3\bar{\Phi}e^{-2\beta(E_i-\mu_i)}+e^{-3\beta(E_i-\mu_i)}]$ and $F_2=\ln[1+3\bar{\Phi} e^{-\beta(E_i+\mu_i)}+3\Phi e^{-2\beta(E_i+\mu_i)}+e^{-3\beta(E_i+\mu_i)}]$. Here the temperature-dependent Polyakov-loop effective potential $\mathcal{U}(\Phi, \bar{\Phi}, T)$ takes the form~\cite{Rob07}
\begin{eqnarray}\label{eq2}
\frac{\mathcal{U}(\Phi,\bar{\Phi},T)}{T^4} &=& -\frac{a(T)}{2} \bar{\Phi} \Phi + b(T) \ln[1 - 6\bar{\Phi} \Phi 
\notag\\
&+& 4(\bar{\Phi}^3 + \Phi^3) - 3(\bar{\Phi} \Phi)^2],
\end{eqnarray}
with
\begin{eqnarray}\label{eq3}
a(T) = a_0 + a_1 \left( \frac{T_0}{T} \right) + a_2 \left( \frac{T_0}{T} \right)^2, b(T) = b_3 \left(\frac{T_0}{T} \right)^3.
\end{eqnarray}
The coefficients $a_0 = 3.51$, $a_1 =-2.47$, $a_2=15.2$ and $b_3=-1.75$ are fitted to pure-gauge lattice QCD thermodynamics~\cite{Rob07}, and $T_0= 210$ MeV is used in the calculation.  In Eq.~\eqref{eq1}, the factor $2N_c$ with $N_c = 3$ accounts for spin and color degeneracy, and $\beta = 1/T$ is the inverse of the temperature. The coupling $G_S$ denotes the scalar interaction strength, while the $K$ term represents the six-point Kobayashi-Maskawa-'t Hooft (KMT) interaction, which breaks the axial $U(1)_A$ symmetry~\cite{Hoo76}. The quark energy for flavor $i$ is $E_i(p)=\sqrt{p^2 +M_i^2}$, where $M_i$ is the constituent quark mass. In the mean-field approximation (MFA), quarks are treated as quasiparticles with constituent masses $M_i$ generated through spontaneous chiral symmetry breaking. The constituent quark mass $M_i$ is determined by the gap equation of
\begin{eqnarray}\label{eq4}
M_i &=& m_i-2G_S\sigma_i+2K\sigma_j\sigma_k,
\end{eqnarray}
where $m_i$ is the current quark mass, $\sigma_i=\langle\bar{q}_iq_i\rangle$ stands for the quark condensate, and ($i$, $j$, $k$) is any permutation of ($u$, $d$, $s$). Following Ref.~\cite{Bub05}, we adopt the parameter set: $m_u$ = $m_d$ =5.5 MeV, $m_s$ = 135.7 MeV, $G_S\Lambda^2$ = 3.67, $K\Lambda^5$ = 9.29, with a momentum integration cutoff $\Lambda$ = 631.4 MeV.

The values of $\sigma_u$, $\sigma_d$, $\sigma_s$, $\Phi$, $\bar{\Phi}$ in the PNJL model are obtained by solving the stationary conditions
{\small
\begin{eqnarray} \label{eq5}
\frac{\partial\Omega_{\textrm{PNJL}}}{\partial\sigma_u}
=\frac{\partial\Omega_{\textrm{PNJL}}}{\partial\sigma_d}
=\frac{\partial\Omega_{\textrm{PNJL}}}{\partial\sigma_s}
=\frac{\partial\Omega_{\textrm{PNJL}}}{\partial\Phi}
=\frac{\partial\Omega_{\textrm{PNJL}}}{\partial\bar{\Phi}}
=0.
\notag\\
\end{eqnarray}}
The pressure, baryon number density, and entropy density follow from standard thermodynamic relations in the grand canonical ensemble:
\begin{eqnarray}\label{eq6}
P=-\Omega_{\textrm{PNJL}}, \quad \rho_B=-\frac{\partial\Omega_{\textrm{PNJL}}}{\partial\mu_B}, \quad s=-\frac{\partial\Omega_{\textrm{PNJL}}}{\partial T},
\end{eqnarray}
and energy density can be calculated as
\begin{eqnarray}\label{eq7}
\varepsilon=-P+Ts+\mu_B\rho_B.
\end{eqnarray}

To study temperature fluctuations, Ref.~\cite{Che25} has developed a systematic theoretical approach by introducing a new thermodynamic function $w = \Omega + Ts$, where $\Omega$ is the thermodynamic potential density, $s$ is the entropy density, and $T$ is the temperature of the system. Its differential equation is given by $dw = Tds - \rho_B d\mu_B$ for a fixed volume $V$, so the temperature can be expressed as $T = \frac{\partial w}{\partial s}$ and the $n$-th order fluctuation of temperature can be written as
\begin{eqnarray}\label{eq8}
\langle (\Delta T)^n \rangle = \frac{\partial^n w}{\partial s^n}, \qquad n\ge 2\ (n\in\mathbb{Z}),
\end{eqnarray}
with $\Delta T=T-\langle T \rangle$ and $\langle\cdots\rangle$ denoting the ensemble average. The dimensionless temperature fluctuation cumulants $c_n$ are defined as
\begin{eqnarray}\label{eq9}
c_n = \frac{\langle (\Delta T)^n \rangle}{T^n},
\end{eqnarray}  
and can be expressed through the $n$-th order derivatives of the pressure with respect to temperature. The first three cumulants correspond to variance, skewness, and kurtosis, expressed respectively as:
\begin{eqnarray}\label{eq10}
c_2 &=& T^2 \left( \frac{\partial^2 P}{\partial T^2} \right)^{-1},
\notag\\
c_3 &=& -T^5 \left( \frac{\partial^2 P}{\partial T^2} \right)^{-3} \frac{\partial^3 P}{\partial T^3},
\notag\\
c_4 &=& T^8 \left[ 3 \left( \frac{\partial^2 P}{\partial T^2} \right)^{-5} \left( \frac{\partial^3 P}{\partial T^3} \right)^2 - \left( \frac{\partial^2 P}{\partial T^2} \right)^{-4} \frac{\partial^4 P}{\partial T^4} \right].
\end{eqnarray} 
Analogous to susceptibilities, the derivatives of pressure with respect to temperature can be extracted and expressed in a dimensionless form
\begin{eqnarray}\label{eq11} 
\chi_n = T^{n-4} \frac{\partial^n p}{\partial T^n}.
\end{eqnarray} 
From the expression for the entropy density in Eq.~\eqref{eq6}, $\chi_1$ and $\chi_2$ are related to entropy and heat capacity, respectively, while higher $\chi_n\ (n\ge2)$ correspond to entropy fluctuations of order $n$. In terms of $\chi_n$, the cumulants of temperature fluctuations become
\begin{eqnarray}\label{eq12} 
c_2=\frac{1}{\chi_2}, \quad \quad c_3=-\frac{\chi_3}{\chi_2^3}, \quad \quad c_4=3\frac{\chi_3^2}{\chi_2^5}-\frac{\chi_4}{\chi_2^4}.
\end{eqnarray} 
Higher-order cumulants, such as $c_5$ and $c_6$, follow as
\begin{eqnarray}\label{eq13} 
c_5 &=& -15\frac{\chi_3^3}{\chi_2^7}+10\frac{\chi_3 \chi_4}{\chi_2^6}-\frac{\chi_5}{\chi_2^5},
\notag\\
c_6 &=& 105\frac{\chi_3^4}{\chi_2^9}-105\frac{\chi_3^2 \chi_4}{\chi_2^8}+10\frac{\chi_4^2}{\chi_2^7}+15\frac{\chi_3 \chi_5}{\chi_2^7}-\frac{\chi_6}{\chi_2^6}.
\end{eqnarray} 
In relativistic heavy-ion collisions, the event-mean transverse momentum $\langle p_T \rangle$ of charged particles scales approximately linearly with an effective temperature, $\langle p_T \rangle \propto T_{\text{eff}}$~\cite{Gar20}. In order to eliminate the influence from this coefficient that is not determined quite well, we instead analyze dimensionless ratios of temperature fluctuation cumulants
\begin{eqnarray}\label{eq14} 
R_{32}=\frac{c_3}{c_2^2}, \quad R_{42}=\frac{c_4}{c_2^3}, \quad R_{52}=\frac{c_5}{c_2^4}, \quad R_{62}=\frac{c_6}{c_2^5}.
\end{eqnarray} 
where the powers of the variance $c_2$ in the denominators are chosen to cancel the powers of $T$ appearing in Eqs.~\eqref{eq9} and \eqref{eq10}.

\begin{figure}[tbh]
\includegraphics[scale=0.55]{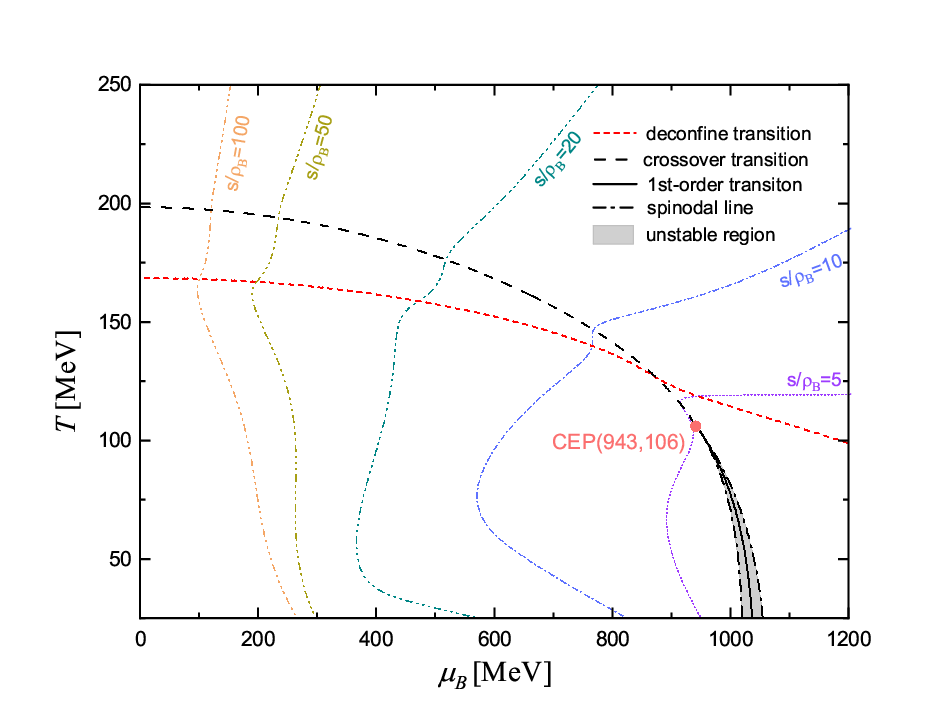}
\caption{(color online) QCD phase diagram in the $\mu_B$-$T$ plane based on the three-flavor PNJL model. The black dashed and solid lines represent the chiral crossover and the first-order phase transition line, respectively. The red dot denotes the critical endpoint (CEP) and the shaded region enclosed by the black dash-dotted lines represents the spinodal unstable region. The red short-dashed line indicates the deconfinement transition line and the different isentropic lines with $s/\rho_B = 5, 10, 20, 50, 100$ are also labeled in the plot.} \label{fig1}
\end{figure}

\section{Results and Discussion}

\begin{figure*}[tbh]
\includegraphics[scale=0.33]{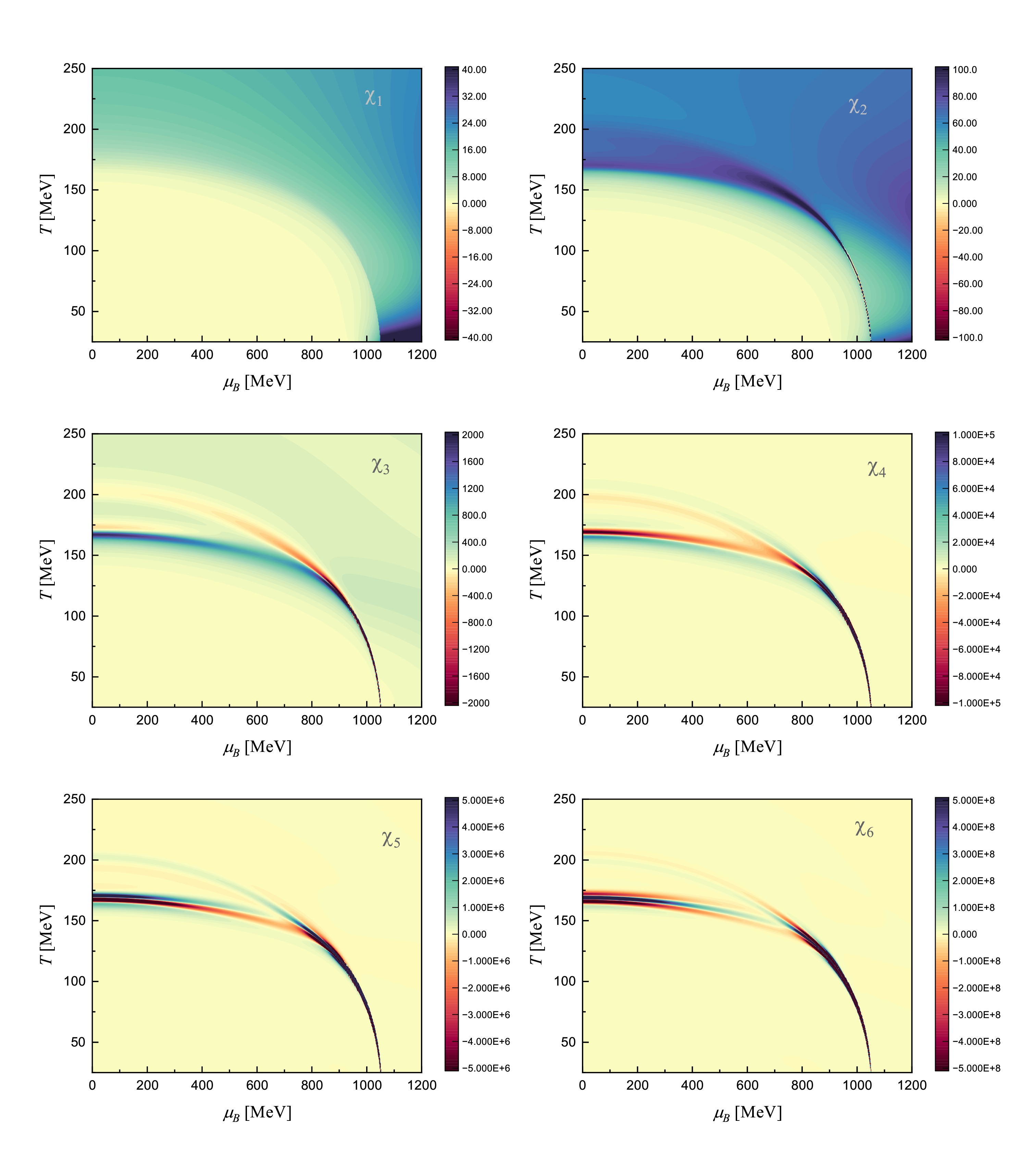}
\caption{(color online) Contour maps of the entropy susceptibilities $\chi_n$ in the $\mu_B$-$T$ plane calculated within the three-flavor PNJL model. Here, $\chi_1$ and $\chi_2$ correspond to the dimensionless entropy and heat capacity, respectively. $\chi_3$ and $\chi_4$ represent the skewness and kurtosis of entropy fluctuations, while $\chi_5$ and $\chi_6$ characterize higher-order fluctuation of the entropy distribution. } \label{fig2}
\end{figure*}

\begin{figure*}[tbh]
\includegraphics[scale=0.33]{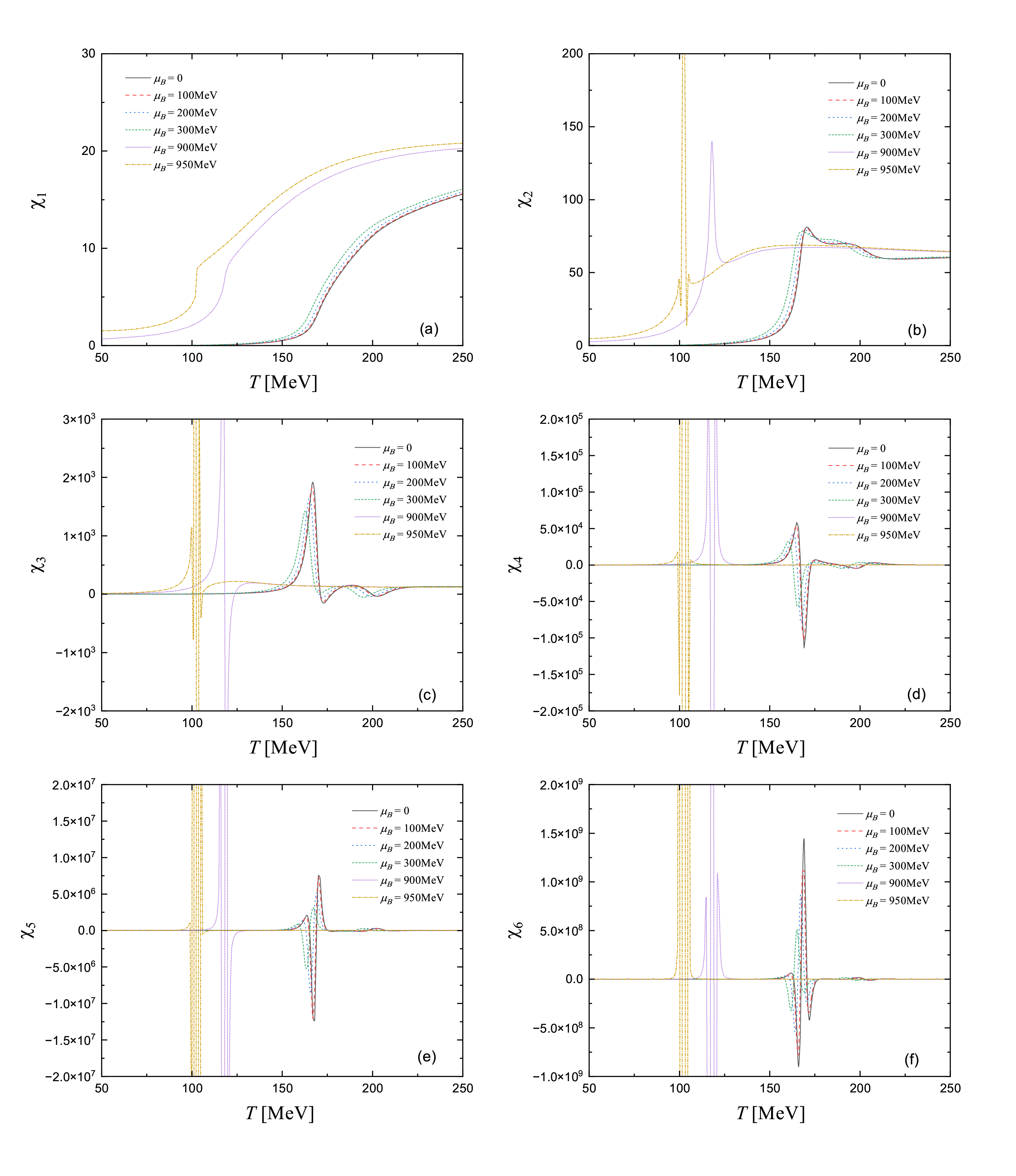}
\caption{(color online) Entropy susceptibilities $\chi_n$ ($n=1$ to $6$) as functions of the temperature at baryon chemical potential $\mu_B$ = 0, 100, 200, 300, 900, and 950 MeV.} \label{fig3}
\end{figure*}

\begin{figure*}[tbh]
\includegraphics[scale=0.33]{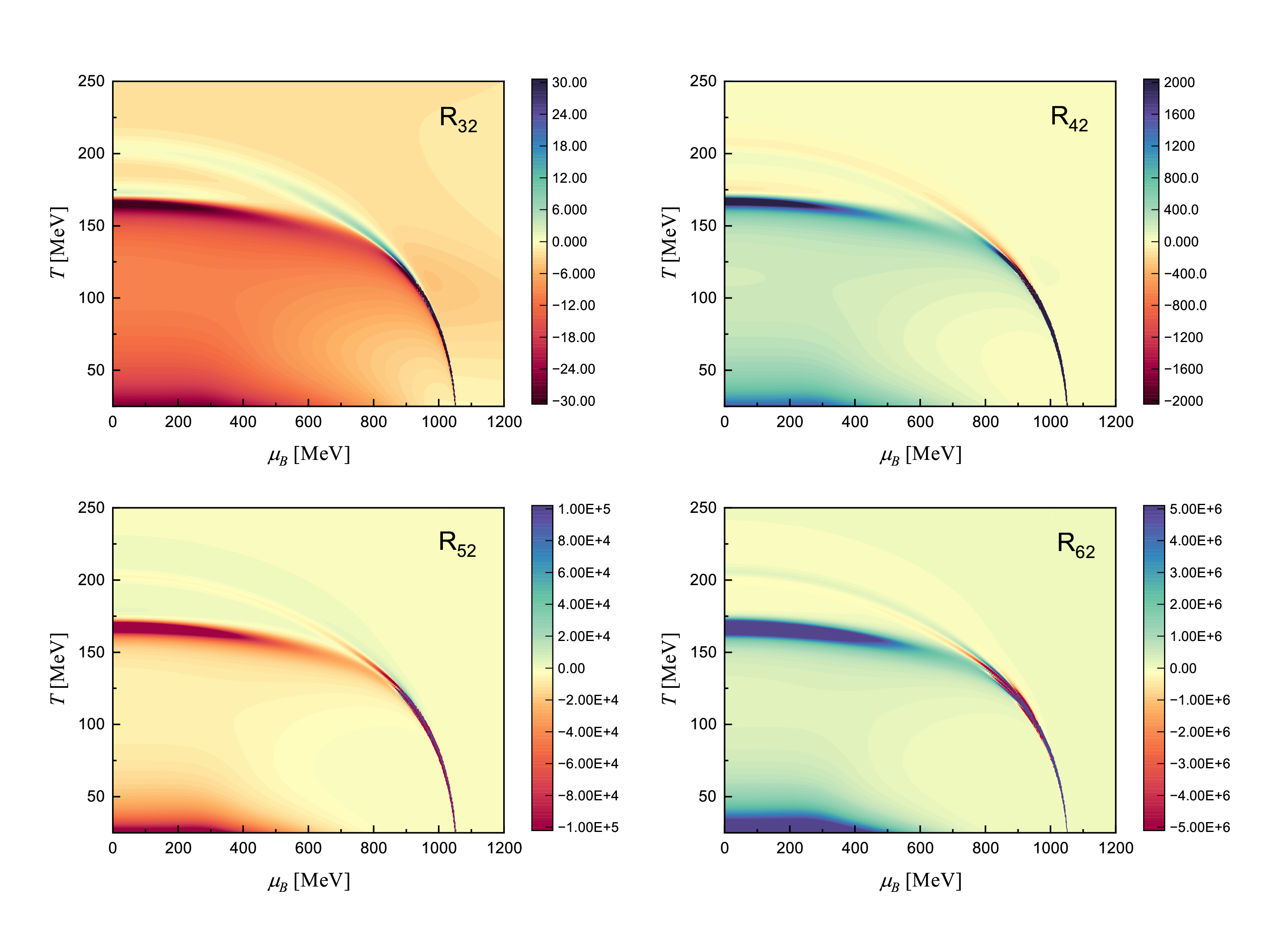}
\caption{(color online) Contour maps of the dimensionless cumulant ratios of temperature fluctuations $R_{32}=c_3/c_2^2$, $R_{42}=c_4/c_2^3$, $R_{52}=c_5/c_2^4$, $R_{62}=c_6/c_2^5$, in the $\mu_B$-$T$ plane based on the three-flavor PNJL model.} \label{fig4}
\end{figure*}

\begin{figure*}[tbh]
\includegraphics[scale=0.33]{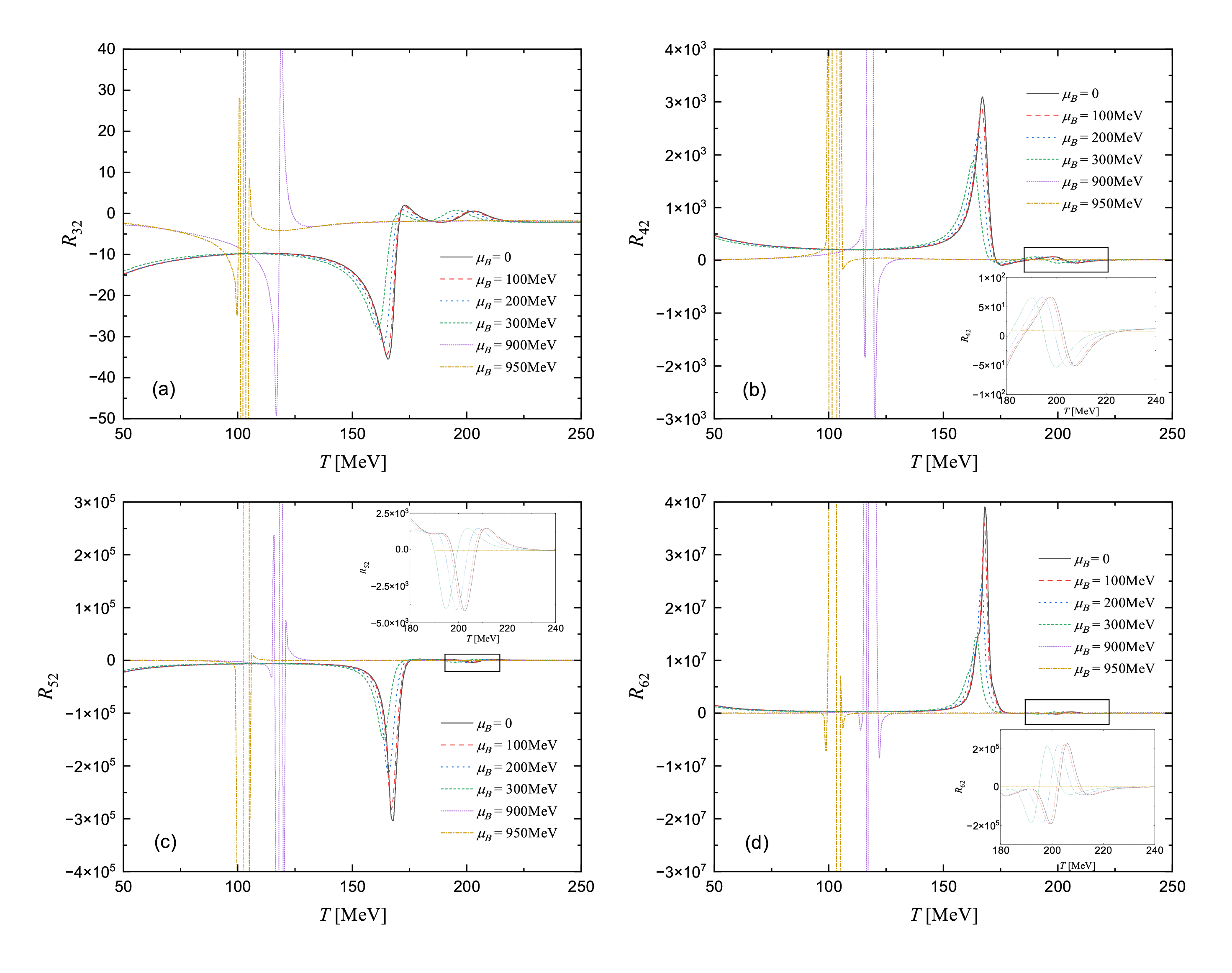}
\caption{(color online) Dimensionless cumulant ratios $R_{n2}$ ($n=3$ to $6$) of temperature fluctuations as functions of the temperature for several values of the baryon chemical potential. Insets provide a close-up view of the crossover region in the temperature range of $180-240$ MeV.} \label{fig5}
\end{figure*}

We first discuss the phase structure in the $\mu_B-T$ plane based on the three-flavor PNJL model. For simplicity, we adopt the symmetric chemical potential setting $\mu_u=\mu_d=\mu_s=\mu_B/3$, where $\mu_B$ is the baryon chemical potential. As shown in Fig.~\ref{fig1}, the black dashed and solid lines correspond to the chiral crossover and first-order phase transition line, respectively, while the red short-dashed line indicates the deconfinement transition. As noted in Ref.~\cite{Sha18}, the chiral crossover and the deconfinement phase transition overlap within a certain region at low chemical potentials. Consequently, we determine these transitions by locating the local maxima of $\partial \langle \bar{u}u \rangle / \partial T$ (chiral crossover) and $\partial \Phi / \partial T$ (deconfinement transition), respectively. The red dot connecting the chiral crossover and first-order phase transition lines marks the critical endpoint (CEP) at $\mu_B = 943$ MeV and $T = 106$ MeV, which can vary depending on the parameter space or types of the models. The shaded region bounded by the black dash-dotted lines is the spinodal (mechanical) instability region and the different isentropic lines with $s/\rho_B = 5, 10, 20, 50, 100$ are also labeled in the plot. The spinodal instability region can be derived by the condition $(\partial P/\partial \rho_B)_T < 0$ in the $\rho_B-T$ plane of the phase diagram, the detailed descriptions can be found in Ref.~\cite{Cos10}.

The entropy susceptibility $\chi_n$ is defined as the $n$-th order derivative of pressure with respect to temperature, normalized by appropriate powers of $T$. This quantity corresponds to the contribution extracted from the expression for temperature fluctuation cumulants, making it a crucial observable in the study of temperature fluctuations. In Fig.~\ref{fig2}, we display the the contour maps of $n$-th order entropy susceptibilities in the full $\mu_B$-$T$ phase diagram, where $\chi_1$ corresponds to the dimensionless entropy ($s/T^3$) and $\chi_2$ represents the dimensionless heat capacity ($C_V/T^3$). Referring to Fig.~\ref{fig1}, the yellow region (where $s/T^3 \approx 0$ and the system is in an ordered state) corresponds to the chiral symmetry breaking phase, while the blue region (where $s/T^3 > 0$ and the system is in a disordered state) represents the restoration of chiral symmetry with vanishing order parameter. The $\chi_2$ contour map more distinctly reveals the first-order phase transition line and the critical endpoint, and the higher-order entropy susceptibilities ($\chi_3$ to $\chi_6$) further highlight entropy fluctuations near the deconfinement transition at low baryon chemical potentials. As shown in Fig.~\ref{fig2}, the closer the entropy fluctuations approach the critical endpoint of the phase transition, the more frequently oscillatory behavior appears. At the meanwhile, the strength and amplitude of the oscillations increase significantly with the order of fluctuations as reflected by the color bars in panels.

To better characterize the variation of entropy susceptibilities across the chiral crossover region, the deconfinement phase transition, and near the critical point, we plot in Fig.~\ref{fig3} the entropy susceptibilities $\chi_n$ ($n=1$ to $6$) as functions of the temperature at baryon chemical potential $\mu_B$ = 0, 100, 200, 300, 900, and 950 MeV. Among the selected constant-$\mu_B$ curves, those with $\mu_B \leq 300$ MeV traverse both the chiral crossover and deconfinement regions at low chemical potentials, whereas the curves at $\mu_B = 900$ MeV and 950 MeV approach the critical region, the latter specifically undergoing a first-order phase transition. In panel (a), $\chi_1$ varies continuously and monotonically with temperature for all curves except at $\mu_B = 950$ MeV, where it exhibits a discontinuous jump at the first-order phase transition. As $\mu_B$ increases, the constant-$\mu_B$ curves shift collectively toward lower temperatures, with the magnitude of this shift becoming more pronounced at higher chemical potentials. In panel (b), all curves exhibit non-monotonic behavior, characterized by a pronounced peak near the critical point. It is worth noting that when crossing the first-order phase transition, two shallow minima appear with slight oscillations. In panels (c) through (f), we can see that the non-monotonic behavior becomes dramatically enhanced for higher-order entropy fluctuations relative to the lower orders, meaning that the higher-order fluctuations are more sensitive to observation in experiments. Similar behaviors of net baryon number fluctuations aroused by the chiral phase transition have been found in the beam energy scan (BES) experiments at RHIC STAR~\cite{Abo23}. Additionally, panels (c) to (f) of Fig.~\ref{fig3} reveal that entropy fluctuations (or oscillations) predominantly occur around three characteristic temperatures: near 200 MeV, corresponding to the chiral crossover; near 170 MeV, associated with the deconfinement phase transition; and near 110 MeV, marking the critical point. As the temperature decreases, the fluctuations in the deconfinement region become stronger than those in the chiral crossover region. When the temperature further approaches the critical point, the fluctuations intensify rapidly and eventually diverge. Given that the entropy density is proportional to the charged particle multiplicity $N_{ch}$, these distinctive features of entropy fluctuations should be observable in heavy-ion collision experiments.

As noted in the Introduction, event-by-event mean transverse momentum fluctuations have been extensively measured across various collision energies and systems, providing a valuable probe into the QCD phase structure. Since the system temperature scales linearly with the event-by-event mean transverse momentum of final-state charged particles, the temperature fluctuations extracted from these measurements offer a novel approach to probe the QCD phase diagram~\cite{App99,Ada03,Adl04,Ach24}. As shown in Fig.~\ref{fig4}, we present the contour maps of the dimensionless cumulant ratios of temperature fluctuations, $R_{n2} = c_n/c_2^{n-1}$ ($n=3,4,5,6$), revealing distinct non-monotonic behaviors across the phase diagram. Consistent with our theoretical findings, these cumulant ratios exhibit slight fluctuation through the chiral crossover region but develop pronounced oscillations around the critical point. In contrast, distinct peak and dip structures are observed around a temperature of 170 MeV at low baryon chemical potential. Specifically, contour maps like $R_{32}$ and $R_{52}$ show dips (red regions), whereas maps such as $R_{42}$ and $R_{62}$ display peaks (blue regions). Fig.~\ref{fig5} further demonstrates these non-monotonic features by plotting $R_{n2}$ versus temperature at fixed baryon chemical potentials ($\mu_B = 0, 100, 200, 300, 900, 950$ MeV). Similar to the observations in Ref.~\cite{Che25}, the peak/dip structures are most prominent at low $\mu_B$ but systematically diminish with increasing $\mu_B$. This behavior arises from a competitive interplay: as $\mu_B$ rises, the sharpening of the chiral phase boundary near the critical point enhances temperature fluctuations, while the amplitude of the deconfinement-related peak/dip structures is progressively suppressed. Consequently, our results indicate that these non-monotonic peak and dip structures in cumulant ratios of temperature fluctuations are associated with the deconfinement phase transition. The distinct peak-dip structures observed at low $\mu_B$ may provide a clear experimental signature for distinguishing deconfinement effects from chiral phase transition signatures in heavy-ion collision data.

\section{Conclusions}
In this study, we investigate temperature fluctuations in hot QCD matter using a three-flavor Polyakov-loop extended Nambu-Jona-Lasinio (PNJL) model. Our analysis establishes explicit connections between temperature fluctuation cumulants of various orders and fundamental thermodynamic quantities, including entropy, heat capacity, and entropy susceptibilities. The results indicate that both higher-order entropy susceptibilities and temperature fluctuation cumulant ratios ($n>2$) exhibit non-monotonic behavior across the chiral phase transition (including crossover, first-order, and critical regions) and near the deconfinement transition. This primarily occurs because the breaking of symmetry during the phase transition leads to a rapid increase or even a sudden change in entropy as the system shifts from an ordered to a disordered state. Consequently, the higher-order susceptibilities of entropy and the corresponding temperature fluctuations display significant non-monotonic variation, with especially pronounced oscillations occurring close to the critical point. We can also see that those non-monotonic behaviors become dramatically enhanced for higher-order entropy fluctuations relative to the lower orders, meaning that the higher-order fluctuations are more sensitive to observation in experiments. A key finding is that temperature fluctuation cumulant ratios ($R_{n2}$) exhibit distinct peak/dip structures at low baryon chemical potentials, consistent with observations in Ref.~\cite{Che25}. We demonstrate that these structures are intrinsically linked to the deconfinement phase transition. As $\mu_B$ increases, the amplitude of these peak/dip features systematically diminishes due to a competitive effect: while temperature fluctuations are enhanced near the critical point (where the chiral phase boundary sharpens), the deconfinement-related fluctuations are progressively suppressed.

In fact, the studies presented in this paper, including the calculations of entropy fluctuations and temperature fluctuations, are all performed under the condition of a fixed chemical potential. The experimental measurement of mean transverse momentum fluctuations is conducted at a fixed multiplicity of charged particles, $N_{ch}$. Since $N_{ch}$ scales directly with the system's entropy ($N_{ch}\sim S$), investigating temperature fluctuations under isentropic (adiabatic) conditions may align better with experimental expectations. Meanwhile, it should be noted that the effect of initial volume fluctuations in low-energy collisions becomes very significant due to the lower multiplicity of charged particles. Furthermore, to improve future theoretical calculations and bring the crossover transition temperature in vacuum closer to lattice QCD simulations or make the chiral phase transition line more consistent with the experimentally predicted chemical freeze-out line, it may be necessary to incorporate more degrees of freedom and interactions. Despite these considerations, the present work provides a qualitative analysis of temperature fluctuation behavior across various phase transition regions. With increased statistics and improved data precision from the RHIC Beam Energy Scan II program, tighter integration between theory and experiment will offer valuable opportunities to probe the phase structure of strongly interacting matter.

\begin{acknowledgments}
We gratefully acknowledge Jinhui Chen for insightful discussions. This work was supported by the National Natural Science Foundation of China (Grant Nos. 12205158, 11975132, 12305148, and 12575134), the Shandong Provincial Natural Science Foundation, China (Grant Nos. ZR2021QA037, ZR2022JQ04, ZR2019YQ01, and ZR2025QC1487), and the Qingdao Natural Science Foundation (Grant No. 25-1-1-4-zyyd-jch).
\end{acknowledgments}

%\bibliographystyle{apsrev4-2} %APS
%\bibliography{ref}

\end{document}